# Prospects for Indirect Dark Matter Searches with the Cherenkov Telescope Array (CTA)

J. Carr [1], C. Balazs [a], T. Bringmann [b], T. Buanes [c], M. K. Daniel [d], M. Doro [e], C. Farnier [f], M. Fornasa [g], J. Gaskins [g], G. A. Gomez-Vargas [h, l], M. Hayashida [i], K. Kohri [j], V. Lefranc [k], A. Morselli [l], E. Moulin [k], N. Mirabal [m], J. Rico [n], T. Saito [o], M.A. Sánchez-Conde [f], M. Wilkinson [p], M. Wood [q], G. Zaharijas [r], H.-S. Zechlin [s].

[1] Aix Marseille Univ., CNRS/IN2P3, CPPM, Marseille, France; [a] Monash University, Clayton, Victoria, Australia; [b] Univ. Oslo, Norway; [c] Univ. Bergen, Norway; [d] Univ. Liverpool, UK; [e] Univ. Padua and INFN Padua, Italy; [f] Stockholm Univ., Sweden; [g] GRAPPA, Univ Amsterdam, Netherlands; [h] Pontifical Catholic Univ., Santiago, Chile; [i] Univ. Tokyo, ICRR, Japan; [j] KEK, Tsukuba, Japan; [k] CEA/IRFU/SPP, CEA-Saclay, France; [l] INFN Roma Tor Vergata, Italy; [m] Univ. Complutense de Madrid, Spain; [n] IFAE, Barcelona, Spain; [o] Univ. Kyoto, Japan; [p] Univ. Leicester, UK; [q] SLAC, USA; [r] Univ. Nova Gorica, Slovenia; [s] Univ. Torino and INFN Torino, Italy.

*For the CTA Consortium*

https://www.cta-observatory.org/

The Cherenkov Telescope Array (CTA) will have a unique chance of discovery for a large range of masses in Weakly Interacting Massive Particles models of dark matter. The principal target for dark matter searches with CTA is the centre of the Galactic Halo. The best strategy is to perform CTA observations within a few degrees of the Galactic Centre, with the Galactic Centre itself and the most intense diffuse emission regions removed from the analysis. Assuming a cuspy dark matter density profile for the Milky Way, 500 hours of observations in this region provide sensitivities to and below the thermal cross-section of dark matter annihilations, for masses between a few hundred GeV and a few tens of TeV; therefore CTA will have a significant chance of discovery in some models. Since the dark matter density in the Milky Way is far from certain in the inner kpc region, other targets are also proposed for observation, like ultra-faint dwarf galaxies such as Segue 1 with 100 hours per year proposed. Beyond these two observational targets, further alternatives, such as Galactic dark clumps, will be considered closer to the actual date of CTA operations. Sensitivity predictions for dark matter searches are given on the various targets taking into account the latest instrument response functions expected for CTA together with a discussion on the systematic uncertainties from the backgrounds.

---

[1]Speaker, *carr@cppm.in2p3.fr*

## 1. Introduction

The existence of dark matter (DM) as the dominant gravitational mass in the Universe is by now well established but the detailed nature of dark matter is at present still unknown and multiple hypotheses endure as to its character. In the form of Weakly Interacting Massive Particles (WIMPs) dark matter particles can self-annihilate, converting their large rest masses into other particles, including gamma rays. Indirect detection from such annihilations provides a unique test of the particle nature of dark matter, in situ in the cosmos. Many reviews exist for dark matter candidates and searches *e. g.* [1] and [2]. The Cherenkov Telescope Array (CTA), which is the next generation gamma-ray observatory with a factor 10 better sensitivity compared to existing facilities as well as many other superior parameters [3], has a unique chance of discovery of dark matter in the form of WIMPs.

In the standard thermal history of the early Universe the WIMP self-annihilation cross-section has a natural value of approximately $3 \times 10^{-26}$ cm$^3$ s$^{-1}$, the "thermal cross-section" [4], which provides the scale for the sensitivity needed to discover dark matter in indirect searches. Particular models for WIMPs such as the neutralino in supersymmetry [5] and the lightest Kaluza-Klein particle of theories with extra dimensions [6], give predictions for gamma-ray energy spectra from the annihilations which are essential ingredients towards the predictions of the sensitivity of the indirect searches. Another vital ingredient is the distribution of dark matter in the targets observed for the search.

This presentation will concentrate on a continuum gamma-ray signature for dark matter. There are also potentially interesting gamma-ray line signatures but a discussion of these signals is beyond the scope of this paper.

## 2. Rate of Gamma-Ray event in Detector

The rate of gamma-rays from dark matter annihilation is usually expressed with a separation of terms which characterise the astrophysical properties of the source and the particle physics contribution to the rate. The astrophysical terms are combined in a "J-factor" as an integral of the squared DM density distribution, ρ, over the line of sight (LOS) and inside the observing angle ΔΩ, with:

$$J(\Delta\Omega) = \int_{\Delta\Omega} d\Omega \int_{LOS} dl \times \rho^2[r(l)]$$

Assuming self-conjugate DM particles, the number of observed events is:

$$N_{DM} = \frac{t_{obs}\, J(\Delta\Omega) <\sigma v>}{8\,\pi\, M_\chi^2} \int_{E_{min}}^{E_{max}} \frac{dN_{DM}}{dE}(E)\, A_{eff}(E)\, dE$$

where:

$t_{obs}$ is the duration time of observation;
$<\sigma v>$ is thermally averaged product self-annihilation cross-section times velocity;
$M_\chi$ is the dark matter particle mass;
$\frac{dN_{DM}}{dE}(E)$ is the energy spectrum of the gammas produced in the annihilation;
$A_{eff}(E)$ is the detector effective area;
$E_{min}$ and $E_{max}$ are the energy limits for the measurement.

For the sensitivity calculations presented in this paper, much of the formulism used, in particular the annihilation spectra come from Cirelli et al. [7].

## 3.   Targets for dark matter searches

The indirect dark matter search with CTA has several possible astrophysical targets each with its own inherent advantages and disadvantages. The Milky Way (MW) galaxy represents a natural place to look for DM signatures and its centre is expected to be the brightest source in the DM induced gamma-ray sky, although the exact magnitude is rather uncertain. The DM halo of the MW should also lead to an annihilation signal observable on large angular scales; however astrophysical galactic foregrounds coupled with the enormous spatial extent and the truly diffuse nature of this galactic DM emission make separation between signal and background challenging. On the other hand, nearby dwarf spheroidal galaxies should provide easier separation of signal and background but comparatively lower signals because of both the distance and lower DM content compared to the Milky Way.

The concordance cosmological Cold Dark Matter model predicts that the formation of visible structures has been guided by gravitational accretion of baryons onto previously formed dark matter over-densities. The astrophysical structures of interest result from the hierarchical formation of DM halos from primordial DM over-densities. Some of the resulting halos could have been sufficiently massive to accrete enough baryons to initiate star formation and give origin to galaxies, including the variety of satellite galaxies we actually observe in the Milky Way halo. In-falling dwarf galaxies (*e.g.,* dwarf irregular galaxies) approaching more central parts of their host halo could have evolved to dwarf spheroidal galaxies (dSphs), *e*.g. [8]. These dwarf galaxies, being highly DM dominated and comparatively close by, form one of the primary targets for CTA observations.

Structure formation predicts gravitationally bound dark matter clumps down to much lower masses than observed for dSph galaxies. The low-mass cut-off of the clump distribution is related to when the DM particles fully decoupled from the heat bath in the early Universe and is expected between $10^{-12}$ $M_{Sun}$ and $10^{-3}$ $M_{Sun}$ for typical WIMP scenarios [9]. The number-count distribution of clumps in MW-like galaxies has been investigated with numerical N-body simulations, *e.g*. [10], [11]. Detailed parameters of the distribution of clumps are very model dependent.

The observational strategy proposed for the CTA Dark Matter programme is focused first on collecting a significant amount of data on the centre of the Galactic Halo. Complementary observations of a dSph galaxy will be conducted to extend the dark matter searches. The Galactic Halo and Large Magellanic Cloud are valuable targets both for dark matter searches and studies of non-thermal processes in astrophysical sources. Data will be searched for continuum emission and line features, and strategies will be adopted according to findings. Clearly discoveries will modify any strategies defined a priori.

### 3.1   Milky Way Galaxy

The centre of the Milky Way has in the past been considered as a target for dark matter searches [12]. More recently, because of the rich field of very high energy (VHE) gamma-ray astrophysical sources in the region, the searches focus on the Galactic halo excluding the central region of galactic latitude b < 0.3° [13]. Even excluding the very central region, the total mass of dark matter in the galactic halo together with its proximity to Earth make it the most promising source for DM searches with CTA. The inconvenience of this target, however, is the fact that being a diffuse source, the integration over the inner halo, while yielding a large signal, gives a very large instrumental background from misidentified charged cosmic rays. Further, there are astrophysical backgrounds from various sources which must be understood, even with the very central region excluded from the analysis. It is believed that the disadvantages of the MW target can be overcome with sufficient experimental effort to control systematic effects in background subtraction or modelling. VHE standard astrophysical processes have steeper spectra than the expected DM-induced gamma-ray continuum emission. Given the wealth of other high energy emitters expected in this region, the search for a DM component requires a very deep exposure to enable detection and detailed spectro-morphological studies; a deep understanding of the instrumental and observational systematics; and accurate measurements of other astrophysical emission in the region to be able to reduce at best contamination to the DM signal. A deep exposure for the Galactic Centre observation will provide the means for an in-depth study and better understanding of the astrophysical emissions in this region.

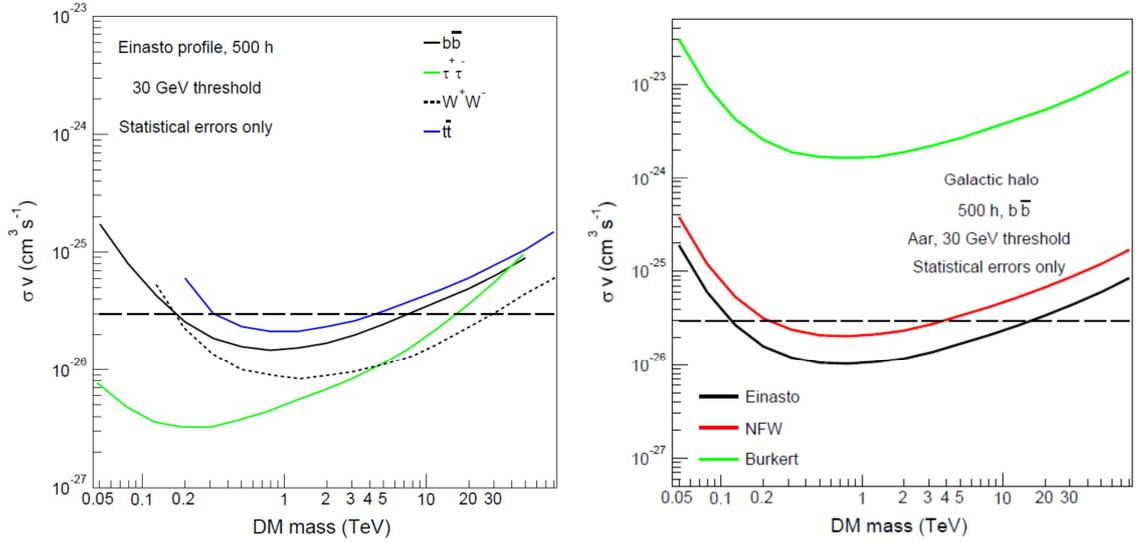

**Figure 1. Left**: Sensitivity for $\sigma v$ from observation on the Galactic Halo with Einsasto dark matter profile and for different annihilation modes as indicated. **Right**: for cuspy (NFW, Einasto) and cored (Burkert) dark matter halo profiles. For both plots only statistical errors are taken into account. The dashed horizontal lines indicate the level of the thermal cross-section of $3 \times 10^{-26}$ cm$^3$ s$^{-1}$.

Stellar dynamics in the Milky Way is dominated by the gravitational potential of baryons up to the kpc scale and the DM density distribution in the inner kpc region can thus be accommodated by both cuspy profiles *e. g.* Navarro, Frenk and White (NFW) [14], Einasto [15] and cored profiles *e. g.* Burkert [16]. Important efforts are ongoing to accurately simulate baryon impact on the DM distribution in the central region of galaxies. With rapid progress being made in the field, a more comprehensive picture for the central region of the Milky Way is expected by the time of CTA observations with reduced theoretical uncertainties on the DM distribution. Although the observation strategy may substantially differ for a kpc-size core profile compared to a cuspy profile, the detection of a DM signal and the detailed study of its morphology would help to resolve this important question.

The Galactic Halo observations will be taken with multiple grid pointings with offsets from the GC position of about ±1.3° to cover the central 4° around the GC as uniformly as possible. This observation strategy defined explicitly to search for DM will require 525 hours to probe cuspy profile DM distributions and the thermal annihilation cross-section. In the CTA Galactic Centre Key Science Programme a further 300 hours are proposed for astrophysics covering up to latitudes 10° from the GC and data are also expected from the extragalactic surveys. These observations should be done in the first three years of CTA operation with high priority.

The sensitivity predictions for observations in the Galactic Halo are shown in Figure 1. The left plot shows the sensitivity for different annihilation modes ($b\bar{b}$, $\tau^+\tau^-$, $W^+W^-$, $t\bar{t}$) and the right plot for various dark matter halo profiles satisfying stellar dynamics as indicated in the caption.

### 3.2  Dwarf Spheroidal Galaxies

The dwarf spheroidal galaxies (dSphs) of the Local Group could give a clear and unambiguous detection of dark matter [17]. They are gravitationally bound objects and are believed to contain up to $O(10^3)$ times more mass in dark matter than in visible matter, making them widely discussed as potential targets. Being small and distant many of the dwarf galaxies will appear as near point sources in CTA and hence the nuisance of the instrumental background is much reduced. Although being less massive than the Milky Way or the LMC, they are also environments with a favourably low astrophysical gamma-ray background making the unambiguous identification of a DM signal easier compared to the Galactic Centre or LMC. Further, the J-factors integrated over the small source have less dependence on the DM profile assumed than the extended sources. Neither astrophysical gamma-ray sources (supernova remnants, pulsar wind nebulae...) nor gas acting as target material for cosmic rays, have been observed in these systems.

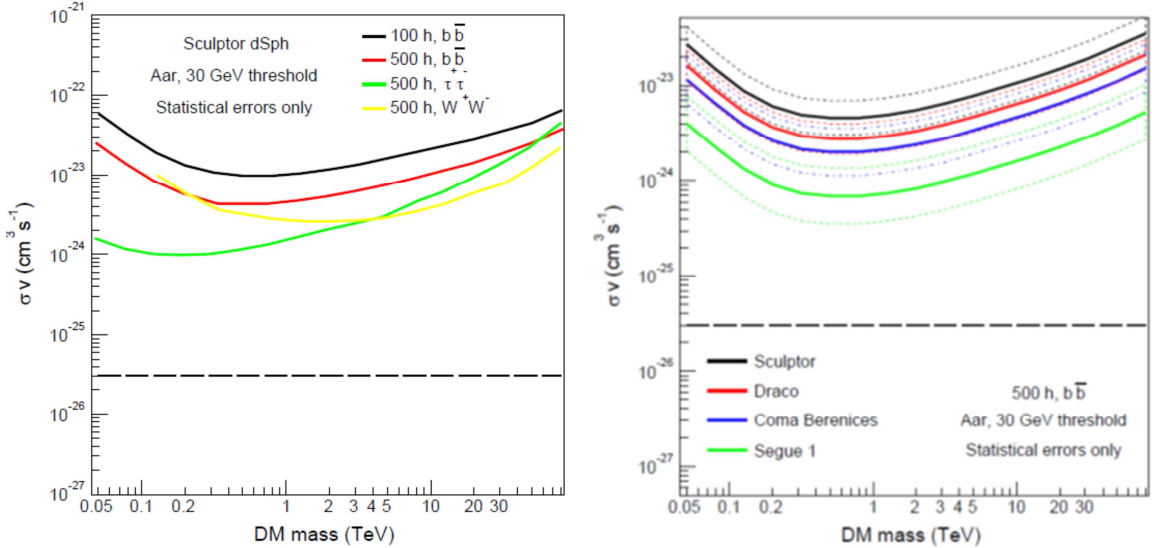

**Figure 2. Left:** Sensitivity for σ v from observation on the classical dwarf galaxy Sculptor for different annihilation modes as indicated. **Right:** Sensitivity for 500 h observation on the classical dSphs Draco and Sculptor, and the ultra-faint dwarf galaxies Segue 1 and Coma Berenices as indicated. Dashed lines correspond to ±1σ on the J-factors. Sensitivity is computed assuming the $b\bar{b}$ annihilation mode and statistical errors only are taken into account.

Due to the larger available sample of spectroscopically measured stars, the classical dwarf galaxies such as Draco, Ursa Minor, Carina, and Fornax have the significantly smaller uncertainties on the J-factor than the ultra-faint dwarf galaxies [17]. However several of the ultra-faint galaxies (*e.g.* Segue 1, Ursa Major II, and Reticulum II) have J-factors which are larger than the J-factors of the best classical dwarfs which to some degree outweighs the larger J-factor uncertainties for these objects. A recent evaluation of dwarf galaxy J-factors was presented in reference [18] which used a Bayesian hierarchical modelling analysis to constrain the DM mass and scale radius in 18 dSph galaxies with good spectroscopic data. Using these mass models we can compare the CTA detection prospects of the dwarf galaxies in this sample. Examples of the sensitivity which could be obtained by observations of a classical dwarf galaxy are in shown in Figure 2 using the same analysis methodology [19] as that performed for the Galactic Halo so the results can be directly compared.

With new data becoming available from recent surveys such as the Dark Energy Survey [20] more nearby dwarf galaxies are being discovered with the exciting possibility that Reticulum II might be a very interesting target for dark matter searches [21]. The final choice of the most promising dwarf galaxy targets for CTA observations will be made at the start of array operation based on all available information at that time.

## 4.  Systematic Errors

The sensitivity estimates presented in Figures 1 and 2 are with only statistical errors. In these figures some indication is given for the uncertainties in dark matter density profiles. The curves for the Galactic Halo with the assumption of cuspy density profiles indicate that the sensitivities can attain the WIMP thermal cross-section and so give hopes of discovery in these models while the curves for dwarf Galaxies and Galactic Halo with cored profiles give sensitivities which are worse by one or two orders of magnitude. These conclusions rest heavily on the control of systematic errors to very low levels.

Much work is in progress to evaluate the systematic errors on the astrophysical J-factors *e. g.* [22].

Many methods exist to evaluate experimental systematic errors. One which is widely used is to introduce nuisance parameters in the likelihood function. The analysis of [23] uses a scaling factor for the predicted signal which is free with a Gaussian distribution for independently each bin in the morphology fit. The true experimental systematic errors are correlated between different bins and so such a method does not easily estimate realistic errors.

Extensive work is needed to design the observational strategy of CTA dark matter searches to understand and minimize the systematic errors.

## 5.  Summary

The existence of dark matter is well established but its nature is still unknown. Of the many possible candidates for dark matter, CTA could have the sensitivity to discover the character of dark matter in models which have been among the most popular over the past decades.

The priority for the CTA dark matter program is to discover the nature of dark matter with a positive observation. The publication of limits following non-observations would certainly happen but in planning the observational strategy the priority of discovery should drive the programme. The possibility of discovery should be considered in the light of the WIMP model predictions where the minimum goal for searches within the most widely considered models is the thermal cross-section of $3 \times 10^{-26}$ cm$^3$ s$^{-1}$.

The sensitivity predictions for the Galactic Halo, the dwarf galaxy Sculptor and the Large Magellanic Cloud are compared in Figure 3. Here it can be clearly seen that the sensitivity possible with the Galactic Halo observation is very much better than that which is possible with a single dwarf galaxy or the LMC. Although this plot does not show the effect of systematics, the relative rankings of the sensitivities is not changed by including them. Figure 4 shows a zoom of the sensitivities in the TeV region compared to pMSSM models from reference [25].

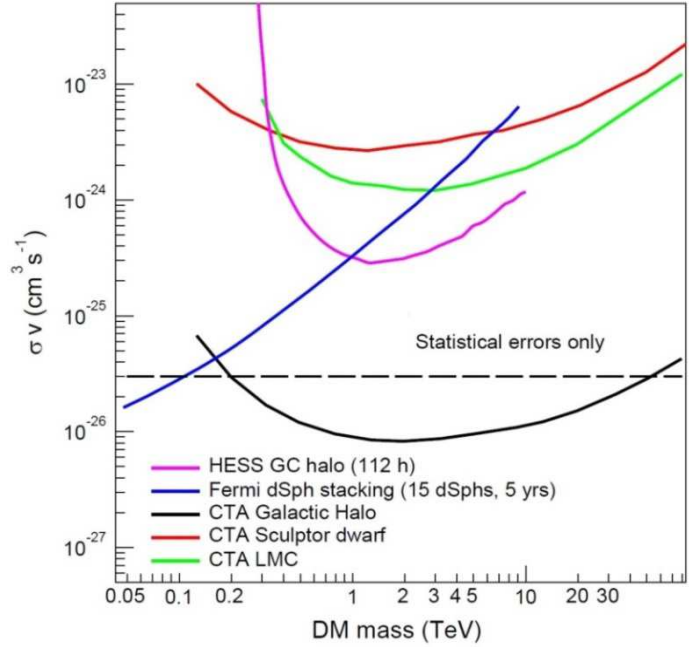

**Figure 3**. Comparison of predicted sensitivities in σ v for the targets of: the Milky Way Halo; the Large Magellanic Cloud and the dwarf galaxy Sculptor. The CTA sensitivity curves use the same method and $W^+W^-$ annihilation modes for each target and the NFW dark matter profile. The sensitivity calculations have a 30 GeV threshold for the MW and Sculptor and 200 GeV for the LMC. The sensitivities for the three targets are all for 500 hours taking into account only statistics errors; for the MW and the LMC, the systematics of backgrounds must be very well controlled to achieve this statistically possible sensitivity. The comparisons with the H.E.S.S. and Fermi-LAT results come from [13] and [24] respectively.

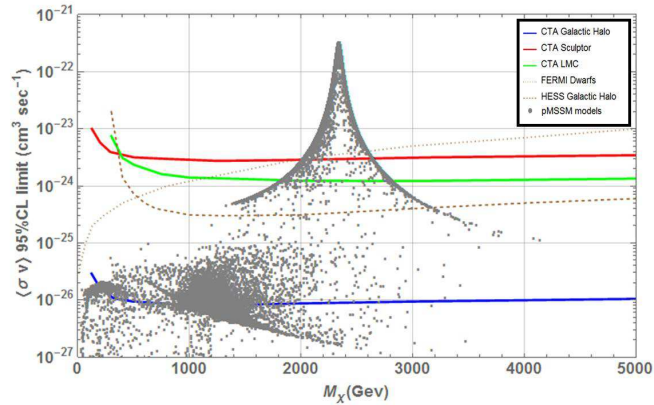

**Figure 4.** Zoomed sensitivities in the TeV mass range together with pMSSM models extracted from [25].

## Acknowledgements

We gratefully acknowledge support from the agencies and organizations listed under Funding Agencies at this website: http://www.cta-observatory.org/.